# A- and B-Exciton Photoluminescence Intensity Ratio as a Measure of Sample Quality for Transition Metal Dichalcogenide Monolayers

*Kathleen M. McCreary,[1]\* Aubrey T. Hanbicki,[1] Saujan V. Sivaram, [1]† Berend T. Jonker[1]*

[1] *Naval Research Laboratory,* Washington DC 20375, USA
† Postdoctoral associate at the Naval Research Laboratory through the National Research Council

\* Author Information:
Correspondence and requests for materials should be addressed to K.M.M. (email: kathleen.mccreary@nrl.navy.mil)

The photoluminescence (PL) in monolayer transition metal dichalcogenides (TMDs) is dominated by recombination of electrons in the conduction band with holes in the spin-orbit split valence bands, and there are two distinct emission features referred to as the A-peak (ground state exciton) and B-peak (higher spin-orbit split state).  The intensity ratio of these two features varies widely and several contradictory interpretations have been reported.  We analyze the room temperature PL from $MoS_2$, $MoSe_2$, $WS_2$, and $WSe_2$ monolayers and show that these variations arise from differences in the non-radiative recombination associated with defect densities. Hence, the relative intensities of the A- and B- emission features can be used to qualitatively asses the non-radiative recombination, and thus the quality of the sample. A low B/A ratio is indicative of low defect density and high sample quality. Emission from TMD monolayers is governed by unique optical selection rules which make them promising materials for valleytronic operations. We observe a notably higher valley polarization in the B-exciton relative to the A-exciton. The high polarization is a consequence of the shorter B-exciton lifetime resulting from rapid relaxation of excitons from the B-exciton to the A-exciton of the valence band.

Monolayer transition metal dichalcogenides (TMDs) are a new class of materials that hold promise for electronic and optoelectronic applications.[1,2] In the past several years, extensive experimental and theoretical research has provided a growing foundation of knowledge for these materials, led by the discovery that many TMDs (*e.g.,* $MoS_2$, $MoSe_2$, $WS_2$, and $WSe_2$) transition from indirect- to direct-gap semiconductors at the monolayer limit.[3,4] Structurally, they are composed of a plane of metal atoms positioned between a top and bottom chalcogen layer, arranged in a hexagonal lattice as viewed normal to the surface (Figure 1 a,b).

Despite rapid progress, there are many fundamental optoelectronic aspects of monolayer TMDs that are not fully understood. The photoluminescence (PL) emission character and interpretation thereof have varied widely. As a case in point: in monolayer $MoS_2$, some works report only a single emission feature associated with optical transitions between the *highest* valence band and the conduction band at the K-points, often referred to as the ground state A-exciton emission.[3,5] However, others observe a strong A-emission feature accompanied by a weaker second emission peak at 100-200 meV higher energy.[6] This second peak, referred to as the B-exciton peak, is associated with transitions between the spin-orbit split valence band and the conduction band. Finally, a number of groups identify two distinct emission peaks of comparable intensities, associated with both A- and B-emissions.[4,7,8]

In addition to disparities in the number of spectral components observed, published interpretations of the relative intensities of these two features are varied and contradictory. The presence of B-peak emission has been associated with high quality samples by some,[9] whereas others claim the opposite.[10] It has been suggested that B-peak emission can only occur in W-based TMDs[11] or when a sample is optically excited below the electronic band gap.[12]

Similar issues exist for the degree of polarization observed in the A- and B-exciton emission from these materials.[13–15] Isolated monolayers have two inequivalent K-valleys at the edges of the Brillouin zone, labeled K and K'. The valence band maxima at K (K') is populated by spin up (spin down) holes, leading to valley dependent optical selection rules[13,14] (Figure 1c). A high degree of valley polarized emission is expected in isolated monolayers.[13] However, experimental values are scattered and typically much lower than predicted.[16–18]

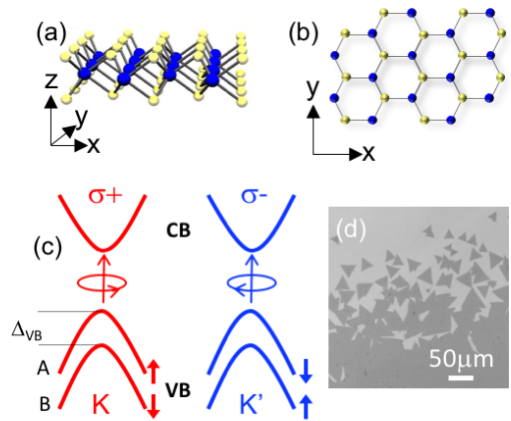

Figure 1: Monolayer transition metal dichalcogenides. (a) The side view and (b) plan view of a monolayer TMD. A plane of metal atoms is positioned between a top and bottom chalcogen layer and arranged in a hexagonal lattice as viewed normal to the surface. (c) A schematic diagram of the single-particle electronic band structure highlighting the spin-split A and B excitonic emission states and the valley dependent optical selection rules. (d) An optical micrograph of a representative monolayer $MoS_2$ sample.

Here we address the discrepancies in interpretation of these PL emission features by analyzing a large number of different monolayer TMDs ($MoS_2$, $MoSe_2$, $WS_2$, $WSe_2$) to better understand the conditions responsible for various emission characteristics and valley polarizations. We find both A- and B-emission intensities can vary widely from sample-to-sample, consistent with other reports, leading to a variety of emission profiles as well as B/A intensity ratios. We show that these observed variations arise from differences in the non-radiative recombination associated with the defect density in a given sample. This relationship between PL profile and exciton dynamics provides a facile method to assess sample quality: a low B/A ratio indicates low defect density and high sample quality, whereas a large B/A ratio signals high defect density and poor-quality material. In comparing the degree of valley polarization from A- and B-excitons for a given sample, we find a notably higher valley polarization in the B-exciton. The high polarization is a consequence of the shorter B-exciton lifetime resulting from rapid relaxation of excitons from the higher energy spin-orbit split state (B-exciton) to the ground state (A-exciton) of the valence band. This additional relaxation pathway shortens the effective lifetime for B-excitons, subsequently reducing the opportunity for intervalley scattering and enhancing the degree of valley polarization.

Results:

Monolayer transition metal dichalcogenides are synthesized via chemical vapor deposition (CVD) on $SiO_2$/Si substrates (275nm oxide) as detailed in the Methods. Figure 1d displays an optical image of $MoS_2$ following a growth. The nucleation density and lateral size of randomly oriented $MoS_2$ flakes varies across the substrate, providing regions of isolated monolayer islands as well as continuous film regions. Triangular growth is typical, although we have also observed hexagonal, star-like, and rounded flakes.

The as-grown TMDs are characterized at room temperature under ambient conditions. Raman, PL, and reflectance measurements are acquired using a commercial system equipped with a 50× objective, so that the area analyzed is approximately a 2 μm diameter circle. The Raman spectra (Figure 2a) of a representative MoS$_2$ sample displays the in-plane $E^1_{2g}$ mode at 384 cm$^{-1}$ and out-of-plane $A_{1g}$ mode at 404 cm$^{-1}$, consistent with monolayer MoS$_2$.[19,20] The normalized differential reflection spectrum (1-R/R$_0$, where R$_0$ is the background signal from the bare substrate) is measured at the same location and displays clear peaks in Figure 2b at 1.86 eV and 2.01 eV, corresponding to the A and B excitonic transitions, respectively. In contrast, only the A peak is easily discernible in the photoluminescence spectra for this particular MoS$_2$ monolayer (Figure 2c), corresponding to radiative recombination via the lowest energy channel. Upon closer inspection and re-plotting the PL intensity on a log-scale (Figure 2d), the B-peak is also evident in PL emission, albeit approximately two orders of magnitude lower than the dominant A-peak.

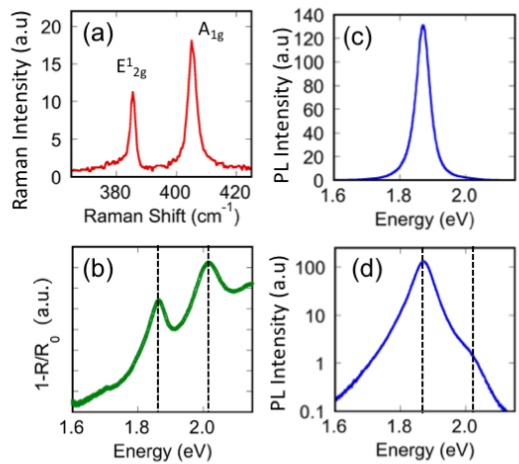

**Figure 2: Characterization of monolayer MoS$_2$.** (a) Raman spectra exhibit in-plane and out-of-plane Raman modes with a peak separation Δk=20 cm$^{-1}$, characteristic of monolayer MoS$_2$. (b) Normalized differential reflection spectra (1-R/R$_0$) find the A and B exciton at 1.86 and 2.01 eV, respectively. Photoluminescence spectra is plotted on (c) linear and (d) logarithmic scale. Emission from both A and B excitons becomes clear when viewed on the log scale. The vertical dashed lines in (b) and (d) are at 1.86 and 2.01 eV. A laser excitation of 532nm is used for Raman and PL measurements.

Significant qualitative differences in the PL emission are reported in the literature for single monolayers, particularly regarding the presence or absence of the B-peak. To gain further insight into these reported differences and to understand the sources and implications of these variations, we examined the PL emission from additional as-grown $MoS_2$ monolayer samples. Figure 3a displays spectra obtained under identical conditions from five different monolayer samples synthesized in three growth runs. The PL intensity of the A exciton varies over an order of magnitude, from 0.113 ct/ms (Figure 3a, red) to 2.315 ct/ms (Figure 3a, light green). Small variations in peak emission energy are also observed, typically within ~10 meV of 1.84 eV, which are likely due to inhomogeneous distributions of strain, doping, and sample-to-substrate distances.[21,22] Figure 3b displays the same spectra as Figure 3a after normalizing to the A-peak intensity and emission energy ($E_A$). It is apparent that the relative contribution of the B-peak changes significantly from sample-to-sample. Furthermore, there appears to be a trend between the intensity of the prominent A-peak and the appearance of a B-peak; samples having lower intensity A-peak emission exhibit a more noticeable B-peak (*e.g.,* Figure 3a,b red curve).

To quantify individual contributions from A- and B- emissions, each PL spectrum is fit with two Lorentzian curves (inset of Figure 3a). In addition to the five samples presented in Figure 3a,b, 19 other MoS$_2$ monolayers, synthesized in 4 separate growth runs, were measured. Interestingly, we observe a non-zero B-peak intensity in all 24 samples. Furthermore, there is a monotonic relationship between the maximum peak intensity for the A- and B-emissions I(*A*) and I(*B*) (Figure 3c) which is well-fit with a simple linear relationship: I(*B*)=0.008+0.009*I(*A*). We note that, surprisingly, the fit does not pass through the origin (0,0), indicating there will be a non-zero B-peak intensity even as the A-emission becomes vanishingly small.

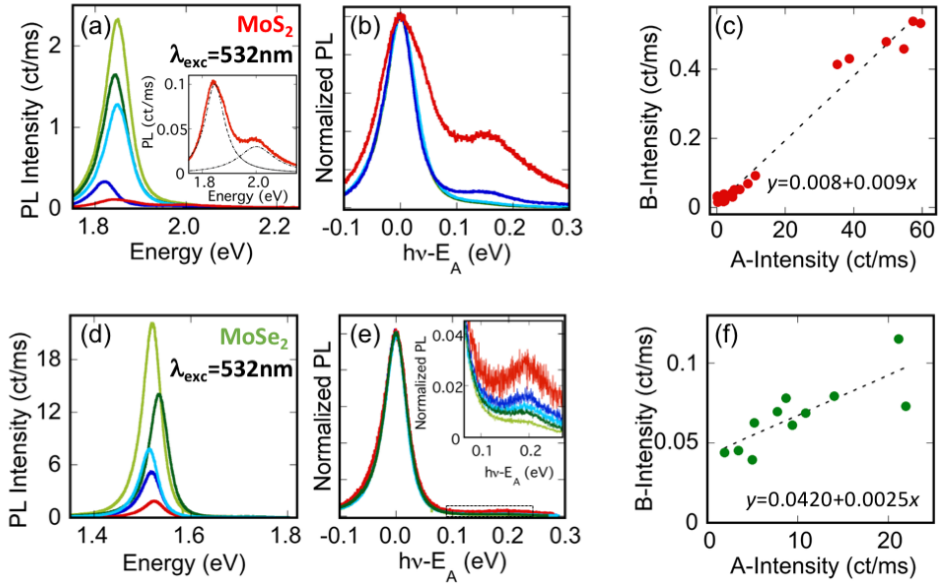

**Figure 3: Photoluminescence from Mo-based monolayer TMDs.** Photoluminescence from representative MoS$_2$ monolayers is plotted in (a) ct/ms and (b) normalized to the maximum emission intensity. All spectra exhibit emission from both A- and B-excitons. The intensity for each peak is determined by 2-Lorentzian fitting, as shown in the inset of (a). (c) The B vs. A intensity is plotted for all 24 MoS$_2$ samples measured, and is well described by the dashed line; y=0.008+0.009x. (d,e) PL from representative MoSe$_2$ is displayed in ct/ms and normalized, respectively. (f) A linear relationship is also observed in MoSe$_2$ B vs. A intensity.

In the majority of samples measured, the B-emission is only a fraction of the A-intensity (<1%) and one could easily overlook the emission from this higher energy feature, particularly when plotting on a linear scale. However, when plotted on a log scale, the B-peak contribution is apparent. For samples exhibiting low-intensity emission from the A-peak, the B-peak becomes especially evident, as shown by the red and dark-blue curves in Figure 3a,b corresponding to samples with the lowest values of I($A$). This analysis indicates that the A and B emission intensities will become comparable in samples that exhibit extremely low intensity A-exciton emission, and that the B-exciton emission will dominate when the A-emission is below 0.008 ct/ms for our measurement conditions.

Photoluminescence measurements were also performed on monolayers of the closely related TMD MoSe$_2$. Emission from the dominant A-exciton occurs at lower energy (relative to MoS$_2$ monolayers) and is observed near 1.52 eV (Figure 3d). Careful inspection reveals a small PL emission peak identified as the B-exciton ~190 meV above the A-exciton (Figure 3e), consistent with the expected valence band splitting and previous observations.[23] Behavior in the selenium-based material is nearly identical to that observed in MoS$_2$ monolayers, with all samples exhibiting a non-zero B-peak emission and a linear relationship between the A- and B-peak intensities with a non-zero intercept, indicating that B-exciton emission persists even as A-emission vanishes (Figure 3f). We note that a similar behavior is observed when using a different laser excitation of 488 nm (2.54 eV). Such excitation is expected to be well above the measured electronic bandgap of ~2.2 eV for MoSe$_2$.[24]

For completeness, we also investigate the tungsten-based TMDs and present the results in Figure 4. In both WS$_2$ and WSe$_2$, the general behaviors observed in MoX$_2$ are repeated; all samples exhibit a measurable B-exciton emission with a linear correlation between A and B emission intensities. Additionally, extrapolation to low emission intensities indicates that the emission from the higher energy B-exciton persists even in the absence of measurable emission from the ground state A-exciton.  We therefore conclude that these are general characteristics for TMD monolayers.

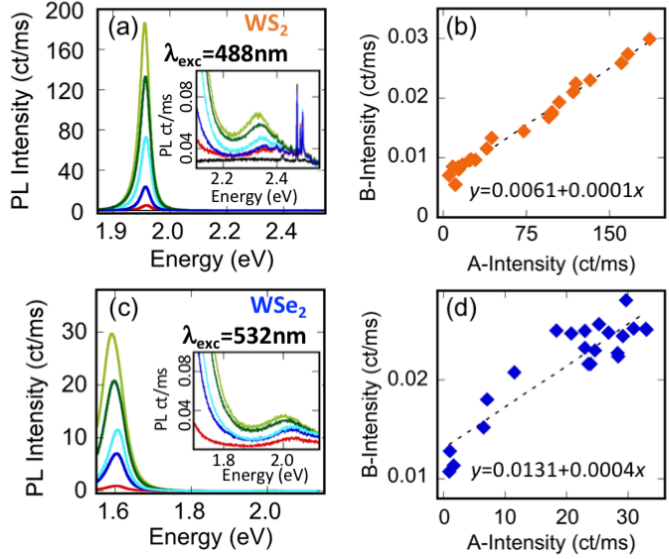

**Figure 4: Photoluminescence from W-based monolayer TMDs.** (a,c) Photoluminescence from $WS_2$ and $WSe_2$ monolayers, respectively. The insets show emission from the B-exciton. For the case of $WS_2$, Raman modes are also evident from $SiO_2$ (black curve) as well as $WS_2$ above 2.4 eV. (b,d) Plots detail the B vs A intensity for $WS_2$ and $WSe_2$ emission, respectively, and find a linear relationship in both instances.

Discussion:

In order to explain the general behavior we observe in our PL data of monolayers, various radiative and non-radiative recombination pathways must be considered. The exciton lifetime ($\tau_E$) is sensitive to both the radiative recombination time ($\tau_R$) and non-radiative recombination time ($\tau_{NR}$) through the relationship

$$\frac{1}{\tau_E} = \frac{1}{\tau_R} + \frac{1}{\tau_{NR}}. \tag{1}$$

This is applicable for the A-exciton lifetime, $\tau_{E,A}$, as well as the B-exciton lifetime, $\tau_{E,B}$. Each material system is investigated using steady-state (cw) excitation conditions which should generate similar initial exciton populations in each sample. Therefore, the sample-to-sample differences in PL intensity ($I_{PL}$) observed for a given monolayer material can be related to variations in the exciton lifetime as $I_{PL} \propto \frac{\tau_E}{\tau_R}$. The radiative recombination time is an intrinsic property, and unlikely to vary at a particular temperature.[25] Non-radiative recombination, however, depends on a variety of factors and can vary widely. In particular, progressively shorter non-radiative recombination times are expected as the defect density increases, providing more non-radiative channels.[26–28] Thus the PL intensity of both A-and B-peaks will be sensitive to changes in the density of defects mediating non-radiative recombination, with the emission intensity decreasing as $\tau_{NR}$ becomes shorter.

Excitons in the B-band have an additional available pathway, in which they scatter to the lower energy A-band. This energetically favored rapid relaxation process, having recombination time $\tau_{B-A}$, is known to occur on the sub-picosecond timescale[29] for $MoS_2$. The additional relaxation pathway will modify equation (1) to

$$\frac{1}{\tau_{E,B}} = \frac{1}{\tau_{R,B}} + \frac{1}{\tau_{NR,B}} + \frac{1}{\tau_{B-A}} \qquad (2)$$

for the B-exciton. While relevant experiments in other TMD monolayers are unavailable, we surmise the associated $\tau_{B-A}$ is comparable to or faster than in $MoS_2$, due to the larger valence band splitting in the other TMDs. This relaxation will reduce the available exciton population in the B-channel while simultaneously increasing the A-band population, resulting in significantly different exciton lifetimes, $\tau_{E,A}$ and $\tau_{E,B}$, and corresponding emission intensities $I_{PL} \propto \frac{\tau_E}{\tau_R}$. Following equation 2, we see that $\tau_{B-A}$ provides an upper bound of ~< 1ps for the B-exciton lifetime, which is significantly shorter than the expected lifetime of ~800ps for the A exciton in $MoS_2$ at room temperature.[25]

In the ideal case, where only radiative recombination and the $\tau_{B-A}$ recombination occur, a significantly higher PL intensity is expected for A-excitons than B-excitons. However, in reality, non-radiative pathways are common, arising from factors such as defects, charge trapping, and exciton-exciton anihiliation,[30–32] and will modify both $\tau_{E,A}$ and $\tau_{E,B}$. The subsequent impact of these additional non-radiative pathways is very different for A- and B-excitons, in that the effect on B-excitons will be minimized due to the very rapid intraband relaxation provided by the $\tau_{B-A}$ pathway. For example, using equation 1 we find that an A-exciton radiative recombination time of 800 ps and non-radiative recombination of 10 ps would result in an effective A- exciton lifetime, $\tau_{E,A}$ of ~10 ps. In contrast, a B-exciton radiative lifetime of 800ps combined with the $\tau_{B-A}$=1ps would first result in an effective B exciton lifetime of ~1ps, which is then only moderately impacted to $\tau_B$~ 900 fs by the addition of the 10ps non-radiative recombination.

Importantly, this indicates that the relative intensities of the A and B exciton emission features of monolayers can be used to qualitatively asses the non-radiative recombination, and thus the quality of the sample. A clearly dominant A-emission indicates a long A-exciton lifetime, relative to the B-exciton lifetime, and a low density of non-radiative defects. In contrast, comparable intensities of A-and B-peaks indicates that non-radiative recombination due to a high density of defects has significantly reduced the effective A-exciton lifetime, such that it is nearly as rapid as that of the B-to-A relaxation pathway, $\tau_{B-A}$. Thus, the B/A peak ratio directly reflects the quality of the sample, with B/A << 1 indicating a low density of non-radiative defects and long A-exciton lifetime. Note that this information can be obtained using cw lasers and steady state conditions, avoiding the more complex apparatus of time-resolved measurement as well as the use of pulsed lasers, which are known to alter the emission spectra and potentially damage the sample due to transient high power densities.[33]

The short lifetime of the B-exciton is ideal for valley polarization. The degree of circularly polarized emission, $P_{circ}$, is related to both exciton lifetime and valley lifetime through the relationship

$$P_{circ} = \frac{P_0}{1+2\frac{\tau_E}{\tau_s}} \qquad (3)$$

where $P_0$ is the initial polarization and $\tau_s$ is the valley relaxation time. As evident in equation (3), for a fixed $P_0$, higher $P_{circ}$ can be obtained either by increasing valley lifetime, or by decreasing exciton lifetime. In a TMD monolayer such as $WS_2$ where the A-emission is dominant, $\tau_{E,B}$ is shorter than $\tau_{E,A}$, as discussed previously. Consequently, emission from the B-exciton should exhibit a higher degree of polarization and provides an internal check of our arguments above.

We test our assertions in monolayer $WS_2$ measured at room temperature. The normalized differential reflection spectrum identifies the A- and B-excitonic features at 1.96 eV and 2.35 eV, respectively (Figure 5a, dashed lines). The degree of valley polarization is measured using 588nm (2.11eV) excitation for the A-peak and 488nm (2.54eV) excitation for the B-peak measurement to be near resonance for each feature, with the excitation energies indicated by the orange and green lines (Figure 5a), respectively. These excitation conditions ensure that the energy separation between laser excitation and PL emission energy is nearly equal for the A- and B-peak measurements, suppressing any dependence on excitation energy.[16,18,26]

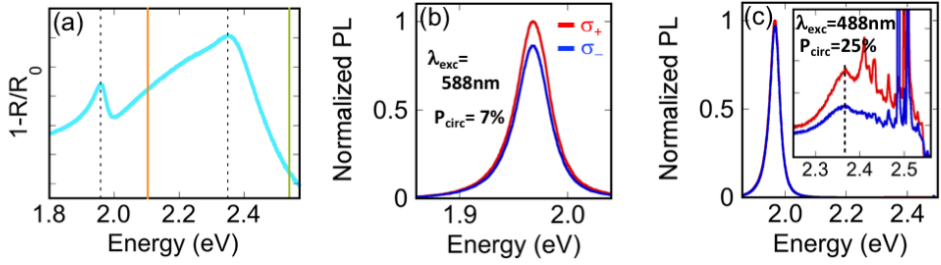

Figure 5: Valley polarization from A- and B-peak emissions in WS$_2$ monolayers. (a) Normalized differential reflection spectrum of a monolayer WS$_2$ region. The dashed lines indicated the A- and B-excitonic features and the colored lines indicate the laser excitation used for polarization measurements of the A-peak (588nm orange line) and B-peak (488nm, green line). (b) σ$_+$ (red) and σ$_-$ (blue) PL components for 588nm excitation. (c) σ$_+$ and σ$_-$ PL components for 488nm excitation. The inset of (c) highlights emission from the B-peak. The narrow peaks above 2.4eV are Raman peaks arising from SiO$_2$ as well as WS$_2$. $P_{circ}$ from the B-peak emission is >3x greater than from the A-peak for the same sample.

The sample is excited with σ+ helicity light, and the emission is analyzed for σ+ and σ- helicity. The subsequent degree of circular polarization is computed as

$$P_{circ} = \frac{I(\sigma_+) - I(\sigma_-)}{I(\sigma_+) + I(\sigma_-)} \qquad (3)$$

where $I(\sigma_{+/-})$ are the polarization resolved PL intensities. As evident in Figure 5b, a $P_{circ}$ of 7% is measured at the A-emission peak for 588nm excitation, comparable to previously reported values under similar conditions.[16,34] However, the circularly polarized emission at the B-peak on the same sample (Figure 5c, inset 2.36eV, dashed line) is significantly enhanced to a value of 25% for 488nm excitation. The narrow peaks present above 2.4 eV arise from Raman features and do not affect the measured valley polarization at the B- emission energy. The considerably higher $P_{circ}$ for B-emission than A-emission can be attributed to the shorter $\tau_{E,B}$ compared to $\tau_{E,A}$ as discussed previously in our model.

Conclusion:

We conclude that the ratio of the B- and A-exciton intensities reflects the density of non-radiative defects, thereby providing a qualitative measure of sample quality. This information helps clarify why significant variations in these PL components have been reported in the literature and enhances our fundamental understanding of excitonic dynamics and valley polarization for both A- and B- emissions in monolayer transition metal dichalcogenides. We identify non-zero PL emission from the B-exciton in all TMD monolayers investigated ($MoS_2$, $MoSe_2$, $WS_2$, and $WSe_2$). Sample-to-sample differences in non-radiative recombination, stemming from differences in defect densities, lead to variations in the A-and B-emission intensities and emission profiles. The variations in the photoluminescence (in the form of B/A intensity ratio) can be utilized to assess sample quality, with a large B/A ratio indicative of high defect density and low sample quality. While all samples exhibit B-exciton emission, the intensity can be orders of magnitude lower than the ground-state A-emission, making identification challenging in high-quality samples. Emission from the B-exciton exhibits an increased degree of valley polarization, relative to the A-exciton. The enhanced polarization results from the presence of an additional relaxation pathway, $\tau_{B\text{-}A}$, which is available only to the B-exciton population and shortens the lifetime $\tau_{E,B}$, subsequently reducing the opportunity for intervalley scattering.

**Methods:**

Separate crucibles of metal trioxide ($WO_3$ or $MoO_3$) and chalcogen (S or Se) serve as the precursors for the monolayer materials $MoS_2$, $WS_2$, $MoSe_2$, and $WSe_2$. A dedicated 2" quartz tube is used for each material to prevent cross contamination, and additional growth details can be found in our previous works.[35,36]

Optical spectroscopy measurements are performed at room temperature in atmosphere using a Horiba LabRam confocal Raman / PL microscope system.

**Acknowledgements**
This research was performed while S.V.S held a National Research Council fellowship at NRL. This work was supported by core programs at NRL and the NRL Nanoscience Institute.